\documentclass[12pt]{article}
\usepackage{epsfig, amssymb, graphicx, amsmath} 
\usepackage{amsthm}
\usepackage{relsize}
\usepackage{comment}
\usepackage{mathcomp}
\usepackage[nottoc]{tocbibind}
\usepackage{fancyhdr}
\usepackage[round-mode=places, round-integer-to-decimal, round-precision=2,
table-format = 1.2, 
table-number-alignment=center,
round-integer-to-decimal,
output-decimal-marker={,}
]{siunitx} 
\usepackage{booktabs}
\usepackage{enumitem}
\usepackage{eurosym}	
\usepackage{wrapfig}	
\usepackage{float}		
\usepackage{footnote}
\usepackage{subcaption}

\usepackage[margin=1in]{geometry}
\usepackage{float}              
\floatstyle{plaintop}           
\restylefloat{table}

\usepackage{booktabs}
\usepackage{dcolumn}
\newcolumntype{d}[1]{D..{#1}}

\usepackage{thmtools}
\usepackage{bbm}		
\usepackage[title]{appendix}
\usepackage{adjustbox}
\usepackage{multirow}
\usepackage{hyperref}
\usepackage{multicol}	
\usepackage[authoryear]{natbib}
\bibpunct{(}{)}{,}{a}{}{,}

\usepackage{xurl}

\numberwithin{equation}{section}

\theoremstyle{definition}

\theoremstyle{remark}

\numberwithin{theorem}{section}
\numberwithin{proposition}{section}
\numberwithin{lemma}{section}
\numberwithin{corollary}{section}
\numberwithin{definition}{section}
\numberwithin{remark}{section}
\numberwithin{example}{section}

\newcommand{\be}{\begin{equation}}
	\newcommand{\en}{\end{equation}}
\newcommand{\ben}{\begin{equation*}}
	\newcommand{\enn}{\end{equation*}}
\newcommand{\bea}{\begin{eqnarray}}
	\newcommand{\ena}{\end{eqnarray}}

\begin{document}
	
	\newlength\tindent
	\setlength{\tindent}{\parindent}
	\setlength{\parindent}{0pt}
	\renewcommand{\indent}{\hspace*{\tindent}}
	
	\begin{savenotes}
		\title{
			\bf{ 
		The puzzle of	Carbon Allowance spread				
		}}
		\author{
			Michele Azzone$^*$, 	Roberto Baviera$^*$ \&  Pietro Manzoni	$^*$	
		}
		
		\maketitle
		
		\vspace*{0.11truein}
		\begin{tabular}{ll}
			$*$ & Politecnico di Milano, Department of Mathematics, Italy.\\			
		\end{tabular}
	\end{savenotes}
	
	\vspace*{0.11truein}
	\begin{abstract}
		\noindent
A growing number of contributions in the literature have identified a puzzle in the European carbon allowance (EUA) market.  Specifically, a persistent cost-of-carry spread (C-spread) over the risk-free rate has been observed. 
We are the first to explain the anomalous C-spread with the credit spread of the corporates involved in the emission trading scheme. We obtain statistical evidence that the C-spread is cointegrated with both this credit spread and the risk-free interest rate. 
This finding has a relevant policy implication: the most effective solution to solve the market anomaly is including the EUA in the list of European Central Bank eligible collateral for refinancing operations. This change in the ECB monetary policy operations would greatly benefit the carbon market and the EU green transition.

	\end{abstract}
	
	\vspace*{0.11truein}

	\vspace*{0.11truein}
	

		\vspace{4cm}

		{\bf Corresponding Author:}\\
		{\bf Roberto Baviera}\\
		Department of Mathematics \\
		Politecnico di Milano\\
		32 p.zza Leonardo da Vinci \\ 
		I-20133 Milano, Italy \\
		Tel. +39-02-2399 4575\\
		roberto.baviera@polimi.it

\newpage
	
	\section{Introduction}

An expanding body of literature is addressing the relationship between carbon spot and futures prices within the European Union Emission Trading Scheme (EU-ETS). Numerous contributions have identified a persistent anomaly in the market-implied cost-of-carry, yet an explanation for its enduring nature remains elusive.
This paper aims to solve this puzzle by conducting a comprehensive analysis focused on the Phase III (2013–2021) of the EU-ETS and aims to analyze the pertinent policy implications.

\bigskip

Set up in 2005, EU-ETS stands as the first multi-country cap and trade system for greenhouse gas (GHG) emissions.
The objective of the system is to  incentivize   companies to reduce their carbon footprint. 
Companies participating  in the scheme  (that are usually called compliance entities or actors) are obliged to submit carbon allowances (hereinafter also EUA) commensurate with their respective $CO_2$ equivalent emissions.
The EU-ETS scheme has been implemented in different phases.


\smallskip

The inaugural phase, Phase I (2005-2007) spanned over  3 years  and is widely considered a pilot phase to prepare for the subsequent one. Notably, in this phase,
almost all allowances were given to companies for free. 
 Pioneering contributions on the Phase I period are \citet{ daskalakis2008european, benz2009modeling,chevallier2009modelling,redl2009price,uhrig2009futures,joyeux2010testing, gorenflo2013futures}.
Phase II (2008-2012) coincided with the first commitment period of the Kyoto Protocol  for the European Union;
throughout this period, the scheme continued to be implemented at the national level, and  the proportion of free allocation over the total allowances was around 90\%. 
Noteworthy contributions addressing  the cost-of-carry puzzle during Phase II include 
	\citet{charles2013market, bredin2016spot, truck2016convenience}; 
the insights gleaned from these studies 
bear direct relevance to our analysis. 

\smallskip

In the first two phases, the EUA spot price experienced frequent collapses, dropping
either to zero or to values close to zero. This can be attributed primarily to the  experimental nature of the EU-ETS framework coupled with the ongoing development of 
 its key legislation (e.g. in April 2011 the European Commission issued the directive that regulated the amount of EUA allocated for free in Phase III\footnote{See \url{https://eur-lex.europa.eu/eli/dec/2011/278/oj} (accessed in December 2023).}). 
Moreover, throughout these two preliminary phases, 
most of the EUA were allocated for free, with the fraction of free EUA consistently exceeding  90\%. 
The resulting  surplus in the volume of issued allowances also contributed in leading the carbon price close to zero for several years across both phases.

\smallskip

Phase III considerably changed the system  compared to previous phases.
The main changes encompassed:
i)	a single, EU-wide, cap on emissions in place of the  national caps of Phases I and II;
ii)	auctioning as the default method for allocating EUA, instead of free allocation;  iii) the obligation for electricity providers to buy all the allowances they need to generate electricity; iv)	a linear decline in the  EU-wide cap by 1.74\% each year, equivalent to 37 million tons of $CO_2$ each year.\footnote{The European Commission estimates that 57\% of the total amount of EUA has been auctioned 
	in Phase III. In Phase IV (2021-2030), the Commission estimates that the share of allowances to be auctioned remains the same. For details see \url{https://climate.ec.europa.eu/eu-action/eu-emissions-trading-system-eu-ets/auctioning_en}   and \url{https://climate.ec.europa.eu/eu-action/eu-emissions-trading-system-eu-ets/free-allocation/allocation-modernise-energy-sector_en} (accessed in December 2023). The linear decline of the cap has been increased to 2.2\% for Phase IV (2021--2030); this is the main change introduced in Phase IV.} This  constraining cap creates a scarcity effect that has positively affected the EUA price.
The substantial  differences with previous phases are the reason why our analysis is focused only on Phase III of the EU-ETS 
scheme.


\bigskip

We analyze the cost-of-carry in the EU-ETS markets in Phase III.
According to financial theory, the cost-of-carry for a financial asset distributing no dividends, such as the EUA, should be a rate reflective of  market player's  financing costs. 
It has been documented in the literature that, starting from Phase II, the  cost-of-carry 
exceeds the risk-free rate \citep{charles2013market, bredin2016spot, truck2016convenience, palao2021inconvenience}: in the following, we call “Carbon Allowance spread” or “C-spread” the cost-of-carry spread over 
the risk-free rate.
The persistence of this  positive spread is the aforementioned cost-of-carry puzzle. The aim of this paper is to solve the puzzle by identifying the financial reason behind this spread. 
We discuss below the contributions in the literature that identify the positive C-spread.

\bigskip
	
	\citet{charles2013market} explore the connection between EU-ETS futures and spot prices in Phase II aiming to  ascertain  whether 
	the cost-of-carry is determined only by the risk-free interest rate.  
	They find evidence of a cointegration relationship among the log-future prices, the log-spot price and the risk-free rate. However, the estimated cointegration coefficients are incompatible with their hypothesis but can be explained considering  the C-spread. 
	They argue that the presence of a spread over the risk-free rate 
is an indicator of inefficiencies within the carbon market.
	\smallskip 
		
	\citet{bredin2016spot} examine the dynamics of the carbon term-structure in the EU-ETS between 2005 and 2014.
They observe that the C-spread is too large and varies too dramatically to be
explained by market interest rates, regardless of the specific rate considered.
	They find statistical evidence that the C-spread of EU-ETS futures is integrated of order 1 starting from Phase II. 
	However, they are unable to establish conclusive evidence regarding a cointegration relationship between the C-spread of the futures with expiry in December (December futures) and the risk-free rate proxied by the yield of government bonds. 
	They suggest that this positive C-spread may be due to financial constraints.
	
	\smallskip
	 \citet{truck2016convenience} study the properties and the determinants of the C-spread.
They analyze the time-series of all December futures contracts active in the Phase II period; i.e.\ they consider seven different time-series of C-spread from December 2008 to December 2014. Their findings suggest a substantial positive C-spread and indicate a negative relationship between C-spread and interest rates, and a positive link with the EUA spot market volatility.

	\smallskip
	More recently, \citet{palao2021inconvenience} apply a quantile regression  to the C-spread of Phase III. In contrast with the previous literature, they consider only the time-series of the front December contract, that they identify as the most liquid one, to build a unique time-series of C-spread; i.e.\ their C-spread time-series, at a given date, is constructed using the future expiring in the front December.
They find statistical evidence of a positive constant term in the regression for  all quantiles of the C-spread; moreover, they observe
a positive correlation with both the EUA realized volatility and some financial market benchmarks. 

\smallskip

In summary,
the C-spread has always remained  positive
since the beginning of Phase II, 
however, there is no consensus in the literature on the reason why this {\it stylized fact} is observed. Previous studies unanimously assert that the C-spread is too large and exhibits excessive variability to be explained by just the movement of the market risk-free rates. 
Our paper solves this C-spread puzzle: we are the first 
to identify the credit spread of the biggest polluting companies in EU as the main cause of the C-spread.
We find statistical evidence that the C-spread and the credit spread are cointegrated throughout Phase III.
To achieve this result, we focus on the time-series of the front December contract, as proposed by \citet{palao2021inconvenience}, and we build an index, the Z-index, as a proxy for the credit spread of  EU-ETS compliance entities.

\bigskip	

The aim of this paper is to investigate the relationship between spot and futures prices negotiated on the EU-ETS markets,  extending the previous studies in three ways: i) we justify an econometric model that considers only the front December contract via a detailed analysis of the trading volumes and of futures roll mechanism, and we also indicate the methodology to construct the Z-index, a proxy of the credit spread of the EU-ETS participants; ii) we find statistical evidence that the C-spread is cointegrated with the Z-index and the risk-free interest rate; identifying the former as its main driver;  iii) we provide empirical findings --by estimating an error correction model--  that the increments of C-spread depend only on the long-run cointegration relationship and not on other financial or commodity drivers.

\smallskip

The rest of the paper is organized as follows. In section \ref{sec:overview}, we indicate the main findings and the key econometric techniques of the paper. In section \ref{sec:dataset}, we describe the dataset and the construction of the C-spread and Z-index time-series. In section \ref{sec:econometric}, we report the results of the econometric analysis, and in section \ref{sect:robu}, we describe the considered robustness checks. Finally, section \ref{sec:conclusions} concludes and closes up with some relevant policy implications.

\section{Overview} \label{sec:overview}
In this section, we explain how to solve   the  C-spread puzzle: we provide the definition of C-spread and delve into the mechanism that determines it. Moreover, we describe the methodology employed to conduct the empirical analysis that supports our conclusions. We claim that the C-spread is driven by the credit spread of EU-ETS market participants.

\smallskip

\begin{figure}[htpb]
	\centering
	\includegraphics[width=0.8\textwidth]{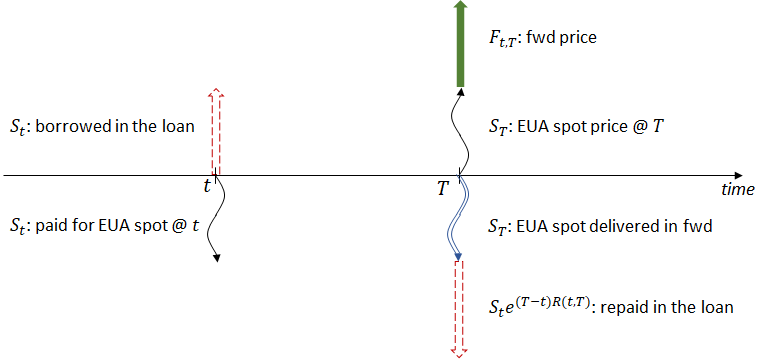} 
	\caption{\small 
Hedging EUA futures. At the value date $t$, the market player sells the forward contract and she will receive the forward price $F_{t,T}$ (straight green arrow) at the expiration date $T$.
In the hedging strategy, she borrows $S_t$ (dashed red arrow) through a loan, to buy the spot EUA contract at the price $S_t$ (wavy arrow). Thus, at maturity $T$, she has the EUA (wavy arrow) that is delivered to the counterparty of the forward contract (double wavy arrow in blue), and she has to pay back the loan, principal amount plus interest rate (dashed red arrow). The interest rate $R(t,T)$ is the one at which the market player gets financing;  thus, the law of one price connect the forward price $F_{t,T}$ with the spot price  $S_t$. Flows are represented according to the usual convention with received flows above the timeline and paid ones below. 
	\label{figure:figura_esempio}}
\end{figure}

In figure \ref{figure:figura_esempio}, we report the static hedging strategy for EUA futures to explain why the credit spread determines the C-spread. In the following description, we focus --for simplicity-- on the forward contracts, because futures and forwards are equivalent in this context.\footnote{This equivalence holds because the EUA spot return and rates’ variations are uncorrelated in our dataset.}
At the value date $t$, the market player sells the forward contract and she will receive the forward price $F_{t, T}$ (straight green arrow) at the expiration date $T$.
In the hedging strategy, she borrows $S_t$ (dashed red arrow) through a loan, to buy the spot EUA contract at the price $S_t$ (wavy arrow). Thus, at maturity $T$, she has the EUA (wavy arrow) that is delivered to the counterparty (double wavy arrow in blue), and she has to pay back the loan with the interest rates (dashed red arrow). 

For the law of one price,  the relationship between the spot and futures prices should depend only on the rate $R(t,T)$, the one at which the market player gets financing. 
This rate is not the risk-free rate, because the participants in the EU-ETS are mainly industrial or electricity generation companies 
whose cost of funding  is equal to the risk-free rate plus a (positive) credit spread.\footnote{We consider only the credit spread of compliance entities and not of regulated credit institutions. We make this choice
because, during the whole Phase III, the capital cost of taking spot or futures EUA positions for these institutions has been considerable: under Basel III standards for market risk, emission allowances
were assimilated to electricity contracts
with	60\%  risk-weight for positions in EUA spot or futures. Thus, they have had the incentive in participating to this market only as intermediaries.  See e.g.,  the Bank of International Settlement (BIS) \textit{Minimum capital requirements for market risk}, §115, p.42, \url{https://www.bis.org/bcbs/publ/d352.htm} (accessed December 2023).  }

\smallskip

This fact justifies the presence of a cost-of-carry spread over the risk-free rate (C-spread).
We define the C-spread at time $t$ as
\begin{equation}
	C_t=\frac{\log \frac{F_{t,T}}{S_t}}{T-t}-r(t,T)\;\;, \label{eq:C_spread}
\end{equation}
where $F_{t,T}$ is the futures price with delivery date  T {in the front December},  $(T-t)$ is the time to maturity of the
futures contract, $S_t$ is the spot price and $r(t,T)$ the risk-free rate between $t$ and $T$.
This definition, from a formal point of view, is equivalent (with the opposite sign) to the definition of the EU-ETS convenience yield in the existing literature \citep[][respectively eq.7 and eq.3]{bredin2016spot,truck2016convenience}.
As previously discussed in the introduction, this C-spread is consistently positive starting from the beginning of Phase II (C-spread puzzle). The main empirical result of this paper is that the C-spread is  equivalent to the credit spread paid by EU-ETS compliance entities.
	\smallskip

Let us explain the financial mechanism behind this result.
A compliance entity that expects to emit GHG in the next years is always short futures EUA.
It can cover its exposure either by buying EUA in the spot market or by entering into  EUA futures contracts. This choice is contingent on the individual entity's credit spread. If the credit spread is above the C-spread  the compliance entity has the incentive to enter into the futures contract; otherwise, it would opt to borrow money at a lower cost and buy the spot EUA.
Furthermore, it is  true that the futures market is an additional potential source of liquidity for a compliance entity that needs funding and holds allowances.\footnote{At the beginning of Phase III, companies in the EU-ETS accumulated a surplus of allowances valid for approximately 1.8 billion tons of $CO_2$, according to \cite{ellerman2014eu}. A survey by \cite{neuhoff2012allocation} suggests that the majority of the surplus EUA is held by the power sector, while a significant portion is held by the industrial sector. Moreover, we remind that EUA allocated for free  are distributed to compliance entities each year on the  $28^{th}$ of February; this allocation occurs more than one year before  the EUA surrendering, see art. 11 of the European Parliament and Council directive 2003/87/EC,  \url{https://eur-lex.europa.eu/legal-content/EN/TXT/?uri=celex\%3A32003L0087} (accessed December 2023).}
The company can address its financing needs by selling the allowances in the spot market and buying futures. This can be an alternative method to borrowing money via traditional methods (loans or  bonds).
Both these mechanisms contribute to the stabilization of the C-spread around the average credit spread in the long-run: changes of the average credit spread  of market participants should be reflected in changes of the C-spread.

\smallskip

In the next section, we introduce an index that serves as a proxy for the average credit spread of an industrial player in the EU-ETS market.
We call this index, which is derived from the Z-spread of the main polluters in the European Union, the Z-index.
One of the primary goals of this  paper is to  provide statistical evidence that the C-spread is explained mostly by the Z-index.
However, this equivalence should not necessarily hold in the short term because of market frictions and  because several companies have different credit spreads and funding needs. But it must be verified in the long-run. In particular, we verify whether the  C-spread is cointegrated with the Z-index during the Phase III.

	\bigskip	

In the realm of time-series analysis, one of the first steps is assessing whether the variables of interest are stationary.
The use of non-stationary data in standard linear regressions leads to \textit{spurious regressions}: ``a regression that looks good under standard measures (...) but which is valueless''\citep[cf.][p.335]{brooks2019introductory}. Moreover, if the data is non-stationary, standard regression hypotheses are not satisfied: even the asymptotic analysis does not hold, and  ``it is therefore not possible to validly undertake hypothesis tests about the regression parameters''\citep[cf.][p.336]{brooks2019introductory}.
Several contributions in the literature point out that the C-spread is integrated of order one in Phase I and II \citep{joyeux2010testing,charles2013market,chang2013mean,gorenflo2013futures}. We  test this property of the  C-spread in the Phase III time-series.
\smallskip

When we conduct an econometric study on a set of  integrated time-series, the established procedure in the literature is to understand whether a cointegration relationship exists to pinpoint one common non-stationary driver. Should such a cointegration relationship be identified, it is also relevant to estimate an error correction model on the first differences of the cointegrated time-series \citep[see e.g.,][chapter 3]{kilian2017structural}, which contains relevant information to the long-run equilibrium. For these reasons, we  utilize the \citet{johansen1988statistical} cointegration tests to examine whether the C-spread and the Z-index are connected in the long-run.

\smallskip

In the existing literature, the work of \citet{bredin2016spot} is the unique contribution that tries to understand whether  a cointegration relationship involving the  C-spread exists. Interestingly, they cannot find conclusive evidence that the C-spread is cointegrated with the risk-free rate in Phase II. We point out that they do not consider the credit spread and, differently from us, they test the cointegration property on the time-series of individual futures instead of considering only the liquid front contract. We solve the C-spread puzzle by finding statistical evidence that the C-spread (of the front contract) is indeed cointegrated with the Z-index in Phase III.

\smallskip

In particular, we are the first to  find statistical evidence that i) the C-spread is integrated of order one in the entire Phase III and ii)  it is cointegrated with the average credit spread  in the EU-ETS market and the  risk-free interest rate; where the former is the main driver of the non-stationary part of the C-spread. 
To obtain this result we have introduced  a new index, the Z-index, of the  average credit spread in the EU-ETS starting from the credit spreads of the individual compliance entities.
Finally, we analyze the first differences of the C-spread with an error correction model and find that the C-spread is connected neither with the variance of the EUA spot nor with global equity and commodity benchmarks.

\smallskip

\citet{bredin2016spot} assert that the C-spread is caused by some financial constraints, yet they are unable to identify them.
We have discovered that the C-spread is predominantly caused by the average credit spread of the EU-ETS compliance entities. This finding has a relevant policy implication:
we claim that the most effective solution to deal  with the market inefficiency identified by the positive C-spread is including the EUA in the list of Euro-system eligible collateral. The inclusion would encourage the formation of an active EUA repo market,  as  market-players would be able to use the spot EUA as collateral for credit operations at the ECB policy rate.\footnote{A repo, or repurchase agreement,  is an agreement to sell a security at a given price and to repurchase it at a pre-specified price at a later date. The eligibility criteria for Euro-system credit operations collateral are specified in  \url{https://www.ecb.europa.eu/pub/pdf/other/gendoc201109en.pdf}, chapter 6 (accessed December 2023). A list of eligible collateral is published and updated daily on the ECB website.}    This  would close the gap between  the policy rate and the repo rate  eliminating the C-spread anomaly.
	Consequently, the   EU-ETS market efficiency would improve, positively affecting the EU green transition.
\section{The dataset} \label{sec:dataset}
In this section, we describe the dataset employed  in the analysis. The study sample is composed of daily data spanning the entire Phase III period; the dataset is divided into three different parts. First,  Euro OIS swaps are employed to bootstrap the risk-free rates' curve. This is the standard methodology for estimating the Euro risk-free rates after the Great Financial Crisis, see e.g., \citet[][]{henrard2014interest}. Secondly, all ICE ECX EUA futures are considered to construct the C-spread  time-series.   Lastly, a  dataset of bonds of the compliance entities in the EU-ETS  is employed to estimate the Z-index. Market data are obtained from Eikon Reuters. In the following, we describe in detail the EU-ETS and the bond dataset, elucidating also the procedure utilized to construct the C-spread and the Z-index. 

\subsection{C-spread dataset}
In this section, we describe the  EU-ETS dataset and explain how we construct the time-series of the C-spread defined in equation \eqref{eq:C_spread} by using the most liquid instruments. Specifically, we consider a front contract time-series of the C-spread selecting the front December futures contract. We utilize the overnight futures as the spot price  and  bootstrap the risk-free rate corresponding to the futures' maturity from the OIS swaps. In the following, we  explain in detail the rationale behind our choices by  analyzing  the trading volumes of EU-ETS contracts.

\smallskip

The EU-ETS dataset consists of the daily closing prices and volumes of overnight futures and futures prices with December maturity, covering the period from January 2013 to December 2020. All futures are negotiated on the ICE market.\footnote{EUA futures also trade on the European Energy Exchange in Leipzig (EEX), however, the vast majority of EUA futures trading takes place on the ICE which is by far more liquid than the EEX.}
The ICE EUA futures contracts have standard characteristics: i) the contract size  is 1000 EUA per lot and the contract price is denominated in Euro per EUA (tick-size 0.01\euro); ii) the expiration date for December EUA futures is the penultimate Monday of the delivery month; iii) all open futures contracts are marked-to-market  with daily margins that are remunerated with the EONIA/\euro-STR 
 rate during Phase III.\footnote{Further details on ICE EUA futures contracts are available at \url{https://www.ice.com/products/197/EUA-Futures} (accessed December 2023).} The EUA market  volume has grown significantly over time: the average yearly volume was half a billion of EUA in Phase I, 5 billion in Phase II, and over 7 billion in Phase III.{\interfootnotelinepenalty10000 \footnote{See \url{https://climate.ec.europa.eu/eu-action/eu-emissions-trading-system-eu-ets/development-eu-ets-2005-2020_en} for the volume up to the end of Phase II and  \url{https://ercst.org/state-of-the-eu-ets-report-2022/} section 8.1.1 for the volume of Phase III (accessed December 2023).} }

\smallskip

\begin{figure}
	\centering
	\includegraphics[width=0.8\textwidth]{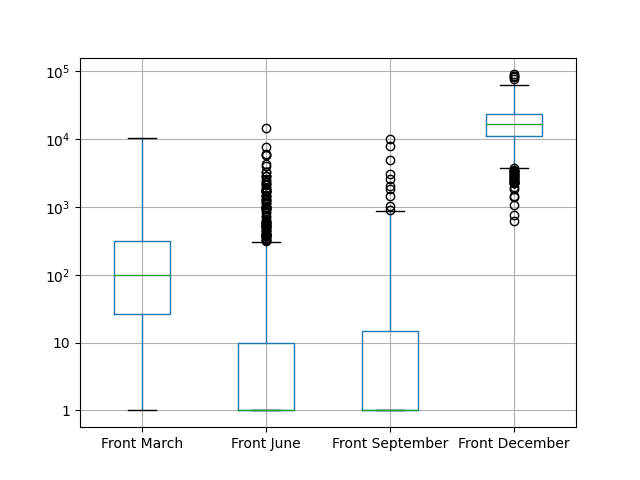} 
	\caption{\small 
		Boxplots of daily volumes in log-scale of the futures expiring in the front March, September and December. The boxplots are computed in the  Phase III  time window.  
	}\label{figure:volume_front}
\end{figure}

To construct a reliable time-series of the C-spread,  we use the most liquid contracts available in the period of interest. In the ICE market,  March, June,  September, and December EUA futures are available. However, we show that March, June, and September futures  are significantly less liquid than the December ones. In figure \ref{figure:volume_front}, we report the boxplots of daily volumes in log-scale of the futures expiring in the front March, June, September, and December in the  Phase III  time window.  The visual inspection of the data reveals  a sharp difference in volumes between the December futures contract and the other three considered contracts: the December futures volume is two orders of magnitude above all the others. 
For this reason, we do not consider March, June and September futures in our analysis. To construct the C-spread we need also the time-series of the EUA spot.  Unfortunately, the European Energy Exchange (EEX) EUA spot is not very liquid: its trading volume is on average two orders of magnitude below the overnight future. 
To avoid this additional source of noise and to accurately measure the implied C-spread, we consider, as in \citet{palao2021inconvenience}, the overnight futures as a reliable  proxy of the spot price.

\smallskip
	
\begin{figure}
	\centering
	\includegraphics[width=0.8\textwidth]{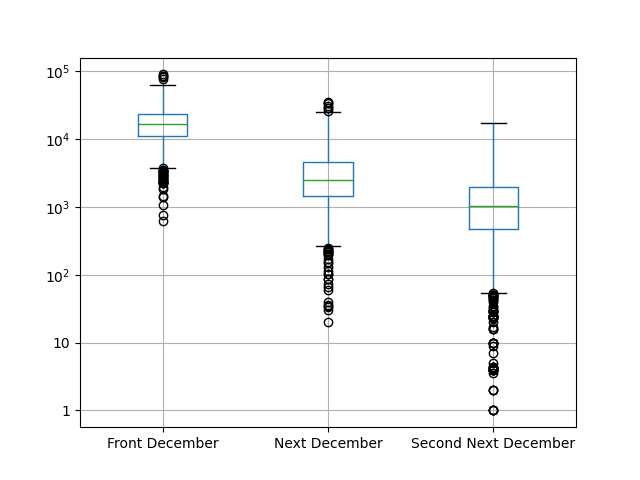} 
	\caption{\small 
		Boxplots of daily volumes in log-scale of the front December contract, the next December contract and the second next December contract. The boxplots are computed in the  Phase III  time window.  
	}\label{figure:volume}
\end{figure}

As already mentioned, we utilize for the analysis a unique time-series of C-spread of the front December contract; again, our choice is motivated by  liquidity considerations. In figure \ref{figure:volume}, we display 		boxplots of daily volumes in log-scale for the  front December contract, the next December contract (i.e.\ the one expiring one year after the front contract), and the second next December contract. The boxplots are computed in the  Phase III  time window. A significant difference in volumes between the front December contract and the next ones, of at least one order of magnitude, is evident. This empirical observation is in line with the established practice in the  market of considering the front December futures as the benchmark for EUA prices.
 Thus, considering a unique time-series of C-spread of the front December futures ensures us consistent results, since we utilize the most liquid futures by several orders of magnitude.
This methodological choice is aligned with the one utilized in \citet{
palao2021inconvenience}. We underline that there is an important distinction between our methodology and previous approaches in the EU-ETS literature wherein the analysis is conducted on all the time-series arising from each individual future.
\smallskip

Moreover, there is a relevant difference between our C-spread time-series and the one in \citet{palao2021inconvenience}, due to the roll mechanism of EUA futures contracts. 
 This rollover is standard in the EUA futures market: participants trade mainly the front December contract, and the operators not interested in entering into a spot EUA position  at maturity rollover their exposure into the next December contract. In this futures market, the rollover starts well over a month before the futures expiry date, as it can be observed by looking at the open interest data.
 For this reason, futures prices present anomalies as they approach their expiry due to the presence of the rollover mechanism as already pointed out by \citet[][p.96]{bredin2016spot}.
  \citet{palao2021inconvenience} exclude several days  from their analysis due to the presence of anomalies in the last month before the December expiry \citep[][footnote 10, p.5]{palao2021inconvenience}. We solve this problem by following a different approach: to deal with this additional source of noise, we switch to the next December contract when the nearest to maturity December futures enters in the last month before the expiry (roll month). In this way, we obtain a unique time-series of C-spread based on very liquid contracts. Descriptive statistics of the C-spread time-series can be found  in section \ref{subsec:desc_stat}.


\subsection{Z-index as a proxy of credit spread}
In this section, we discuss the construction of  our proxy of the average credit spread in the EU-ETS market: the credit spread index (Z-index). 
In particular, we construct the Z-index from the Z-spreads of a dataset of  bonds issued by the European main polluters among the EU-ETS compliance entities. The Z-spread is the spread that would need to be added to the risk-free
zero rate curve so that the discounted cash flows of the  bond are equal to its current market price \citep[see e.g.,][]{choudhry2005understanding}; it is a standard measure in the bond market to estimate the credit spread over the risk-free rate that an issuer has to pay to borrow money.
 We aim  to estimate the average credit spread in this market, at a given day, by computing a weighted average of the Z-spreads of the aforementioned bonds.

\smallskip

In this analysis, we consider the 12 European issuers that are included in the scheme with the highest Scope 1  carbon equivalent emissions. Scope 1 covers emissions from sources that an organization owns or controls directly. Amongst compliance entities, larger companies and larger emitters are more likely to participate in the carbon market \citep{abrell2022corporate}. A possible explanation is that they already have an established trading infrastructure in place and this makes  easier and cheaper for them to get involved in the carbon market \citep{meyer2022emission}. In table \ref{tab:main_poll}, we report the issuers' names, their average yearly carbon emission in the Phase III period (in million tons of $CO_2$),  and their  sector according to the Industry Classification Benchmark (ICB). We consider, in the construction of the Z-index, only the main polluters and no regulated credit institutions because the capital cost of entering a spot or futures EUA position for these institutions has been considerable in Phase III.

\begin{table}
	\begin{center}
		\begin{tabular}{ l| l c l }  

Issuer			&	Ticker &  $CO_2$ Emissions &ICB Sector\\
\hline
		Arcelor Mittal & MT.AS & 169& Industrial Metals and Mining\\
			ENEL & ENEI.MI &96& Electricity\\ 
				ENGIE & ENGIE.PA & 95& Gas Water and Multi-utilities\\
							Lafarge & Unlisted & 92& Construction and Materials\\
								Heidelberg Materials &       HEIG.DE & 64&Construction and Materials\\
																														EDF & Unlisted & 57&Electricity\\
											ENI & ENI.MI & 42& Oil Gas and Coal\\

	Total Energies & TTEF.PA & 41&Oil Gas and Coal\\

			E ON & EONGn.DE & 40&Gas Water and Multi-utilities\\
		AP Moeller & TTEF.PA & 35& Industrial Trasportation\\
					CEZ & CEZ.PR & 28&Electricity\\
	Veolia Enviroment & VIE.PA & 28&Gas Water and Multi-utilities\\
		\end{tabular}
		\caption{\small Issuers considered in the analysis, their average yearly carbon emission in the Phase III period (in million tons of $CO_2$)  and their ICB sector. We consider the first twelve bond issuers by Scope 1 carbon equivalent emissions in the European Union. \label{tab:main_poll}}
	\end{center}
\end{table}
\smallskip

For each  polluter, we select all issued bonds with fixed coupon, non-convertible,  denominated in euro,  of benchmark size (with notional of 500 million or above), traded at least one day in the Phase III period. 
The construction of the daily Z-index follows three steps. First, for each bond, we compute the Z-spread. 
Secondly, we compute the Z-spread of the individual issuer as the average of the Z-spreads of all its issued bonds (that traded on that day) weighted by the bond issued amounts. Lastly, we compute the Z-index ($Z_t)$ by taking the average of the issuers' Z-spread.\footnote{In section \ref{sect:robu}, we repeat all our econometric analyses considering different definitions of the Z-index  to show that our results are robust to the selection of  the proxy for the credit spread.}

\subsection{Summary statistics}
\label{subsec:desc_stat}
In this section, we present some descriptive statistics for  the three key time-series of our analysis: the C-spread $C_t$, the Z-index $Z_t$, and the three month risk-free rate $r_t$. 

\smallskip

\begin{figure}
\centering
\includegraphics[width=0.8\textwidth]{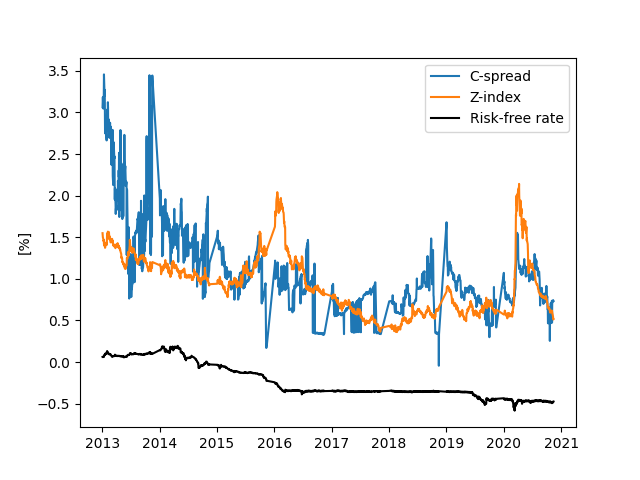} 
\caption{\small 
	C-spread, Z-index, and three-month rate $r_t$ for Phase III in percentage (same scale for all time-series). The C-spread and the Z-index have the same order of magnitude (while the rate is significantly smaller) and appear to have a strong long-run relationship.
}\label{figure:vol}
\end{figure}

In figure \ref{figure:vol}, we plot the three time-series  in percentage (same scale for all time-series) during the Phase III time window. Interestingly, the C-spread and the Z-index have the same order of magnitude (while the rate is significantly smaller) and appear to have a strong long-run relationship. In the next section, we test whether there is statistical evidence of a cointegration relationship between these time-series.
Table \ref{tab:descriptive} presents summary statistics of the three time-series in the period of interest. We observe that the C-spread and the Z-index have similar sample averages and standard deviations.

\begin{table}
	\begin{center}
		\begin{tabular}{ c|c c c }  
			&	$C_t$& $Z_t$ &$r_t$   \\
			\hline
			Average	&$1.1\%$& $1.0\%$ & $-0.2\%$  \\
			Standard deviation &$0.5\%$& $0.4\%$&$0.2\%$
		\end{tabular}
		\caption{ \small Descriptive statistics of the time-series in the Phase III period.\label{tab:descriptive}}
	\end{center}
\end{table}

\smallskip

\begin{figure}
	\begin{subfigure}{.5\linewidth}
		\centering
		\includegraphics[width=8cm]{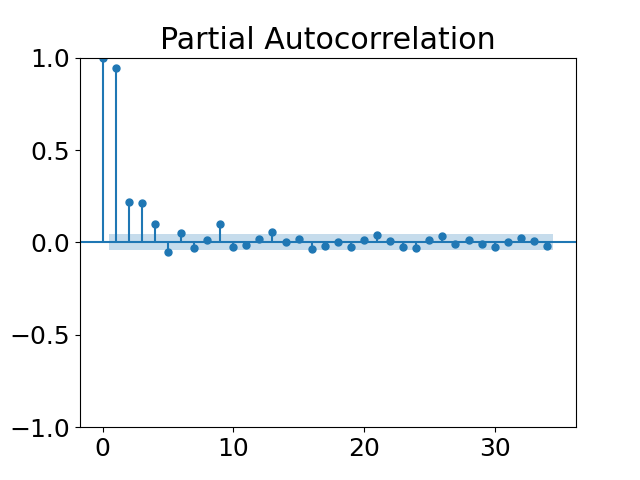}
		\caption{\small  C-spread}
		\label{fig:sub1}
	\end{subfigure}%
	\begin{subfigure}{.5\linewidth}
		\centering
		\includegraphics[width=8cm]{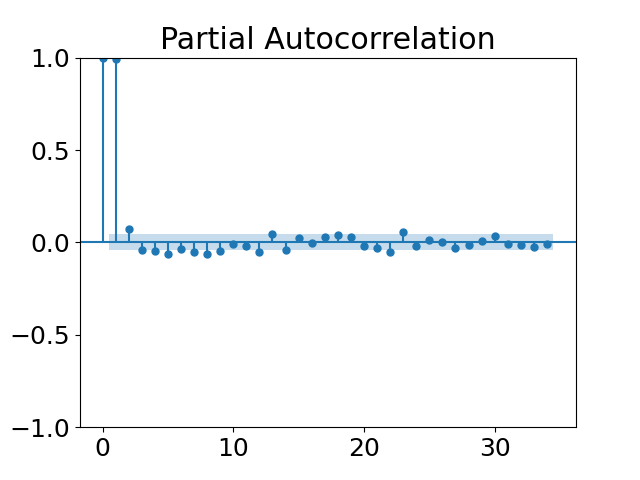}
		\caption{\small Z-index}
		\label{fig:sub2}
	\end{subfigure}\\[1ex]
	\begin{subfigure}{\linewidth}
		\centering
		\includegraphics[width=8cm]{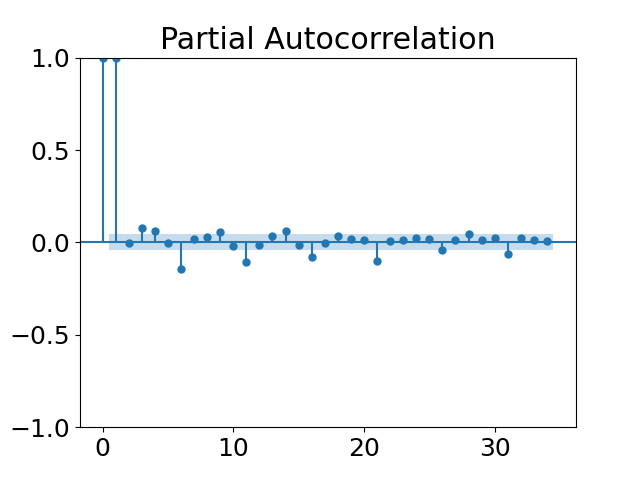}
		\caption{\small  Risk-free rate}
		\label{fig:sub3}
	\end{subfigure}
	\caption{\small Partial autocorrelation function for the three time-series and 5\% rejection region for the null hypothesis of no partial autocorrelation. The first lag of the autocorrelation appears very close to one.}
	\label{fig:parcorr}
\end{figure}

In figure \ref{fig:parcorr}, we plot the partial autocorrelation  function for the three time-series. In all cases, the first lag of the autocorrelation appears very close to one. An autocorrelation close to one is an indicator that a time-series may be non-stationary. As already discussed, there is clear evidence in the literature that the C-spread is non-stationary (integrated of order 1) in the Phase II period. In the next section, we provide statistical evidence that all three time-series, $C_t$, $Z_t$ and $r_t$, are integrated of order 1 in Phase III.

\section{Econometric analysis} \label{sec:econometric}
In this section, we present the main findings derived from the empirical analysis of the C-spread. Our claim is that the C-spread observed in the EU-ETS market is mainly driven (in the long-run) by the credit spread of the compliance entities. We also include in our analysis the risk-free rate, as already suggested by \citet{bredin2016spot}. We find statistical evidence indicating that the C-spread exhibits first-order integration and is cointegrated with both the Z-index and the risk-free rate. Through the estimation of an error correction model, we ascertain that the first differences of the C-spread have an autoregressive structure and that the spread tends to revert to the equilibrium of the cointegration relationship. Finally, our analysis   finds no evidence that the C-spread is correlated with the EUA spot volatility and with other financial asset classes.

\smallskip 

 In particular,  we expect that the connection between the C-spread and the Z-index is a long-run, rather than a short-run, phenomenon. In the short-run, there might be deviations in the C-spread that can be induced by, for example, thin trading or lags in information transmission in the spot price and futures price markets \citep{maslyuk2009cointegration}.  

\smallskip

Before testing for cointegration, non-stationarity must be established.
 We apply the Augmented Dickey-Fuller Generalized Least Squares (ADF GLS)  test  \citep{elliott1992efficient} to verify the null hypothesis of non-stationarity on the C-spread, the Z-index, and the risk-free rate and find that all the time-series have a unit root. When the same test is applied to the first differences of the series ($\Delta C_t$, $\Delta Z_t$ and $\Delta r_t$), these are found to be stationary. In other words, we have statistical evidence that $C_t$, $Z_t$, and $r_t$ are integrated of order 1 in the whole Phase III. This result is in accordance with the  partial autocorrelation functions in figure \ref{fig:parcorr}. As already pointed out, due to the non-stationarity of the C-spread, a standard regression analysis is not viable and yields  meaningless results from a statistical point of view.  Consequently, we verify whether a cointegration relationship between the three time-series exists.

\smallskip

\begin{table}
	\begin{center}
		\begin{tabular}{ c| c c }  
			
			Time-Series	&\multicolumn{2}{c}{ADF-GLS}\\
			&	P-value & Test Statistic  \\
			\hline
			$C_t$& 48\% & -0.58 \\ 
			$\Delta C_t$ & $<0.1\%$ & -7.30    \\ 
						$Z_t$  & 34\% & -0.90\\
			$\Delta Z_t$ & $<0.1\%$ & -9.10 \\ 
			$r_t$& 93\% & 1.10 \\ 
			$\Delta r_t$ & $<0.1\%$ & -8.99  \\ 
		\end{tabular}
		\caption{\small  P-value and test statistic of an ADF GLS  test for the C-spread, the 	Z-index, the risk-free rate, and their first differences ($\Delta C_t$, $\Delta Z_t$ and $\Delta r_t$). We accept the null hypotheses of a unit root for the time-series and reject the null hypotheses of a unit root for their differences. We conclude that the time-series are integrated of order 1.}
	\end{center}
\end{table}

\citet{johansen1988statistical} introduces a procedure for identifying  cointegration relationships within a set  of $g\geq2$  variables integrated of order one (in our case, we have three integrated time-series, hence, $g=3$). The method centers around the determination of the number of  cointegration relationships $q$, which is such that $0\leq q<g$. In particular,
\citet{johansen1991estimation} proposes two statistical tests with the same null hypothesis that are called  trace and  eigenvalues tests. The trace tests  are conducted in a sequence until the number of cointegration relationships $q$ is established (the same procedure is followed for the eigenvalues tests). The first test has null hypothesis  $q\leq0$.  If the null is rejected the second test has null hypothesis  $q\leq1$, and so forth until the null is accepted.
The results of the cointegration tests are presented in table \ref{tab:coint}. The null hypothesis of no cointegrating vector ($q=0$) is rejected  at the 1\% significance level (cf.\ row 1 of the table). Instead, we accept the null of one cointegrating relationship $(q=1)$ as indicated by both trace and eigenvalues tests (cf.\ row 2 of the table).
\begin{table}
	\begin{center}
		\begin{tabular}{| c l l |ccc| ccc|}  
			\hline
			Null	& Trace Stat. & Eig Stat.&  \multicolumn{3}{c|}{Trace Critical Values}&\multicolumn{3}{c|}{Eig Critical Values}  \\
			&&&90\%&95\%&99\%&99\%&95\%&99\%\\
			\hline
			$q\leq 0$& {51.0***}&{46.1***}&21.8&24.3&29.5&15.7&17.8&22.3\\
			$\mathbf{q\leq 1}$&5.0&5.0&9.1&12.3&16.4&9.5&11.2&15.1\\
			$q\leq 2$& 0.0&0.0&3.0&4.1&6.9&3.0&4.1&6.9\\
			\hline
		\end{tabular}
		
		\caption{\small Trace and eigenvalues statistics and critical values of the Johansen test  for different numbers of cointegration relationship $q$. For both tests, we reject the null hypothesis of no cointegration relation and accept the null hypothesis of $q\leq 1$. *, **, *** indicate statistical significance at the $10\%, 5\%, 1\%$ significance levels respectively. \label{tab:coint}}
	\end{center}
\end{table}
The   cointegration vector   of $C_t$, $Z_t$ and $r_t$  estimated with the Johansen procedure is 
\begin{equation}
\{1,-1.21,-0.40 \}\;\;. \label{eq:coint}
\end{equation}

\smallskip 

This relationship is one of the main empirical  results of the paper and it is worth to discuss it in detail. First, we have statistical evidence of only one cointegration relationship ($q=1$): the non-stationary term in the C-spread is explained by the linear combination $$\gamma_1\,Z_t+\gamma_2\,r_t:= 1.21\,Z_t+0.4\,r_t\;\;.$$ This result  verifies the theoretical connection between the C-spread and the Z-index in the long-run. Secondly, the impact of the risk-free rate is moderate. Multiplying the components of the cointegration vector with the averages of the time-series (cf.\ table \ref{tab:descriptive}), we observe that $r_t$ accounts approximately for 6\% of the non-stationary term in $C_t$. Consequently, we can deduce that the Z-index is the main driver of the C-spread. We also point out that the cointegration between the credit spread and the risk-free rate is expected and is well established in the literature \citep[see e.g.,][]{morris1998credit}.

\smallskip

\begin{figure}
	\centering
	\includegraphics[width=0.8\textwidth]{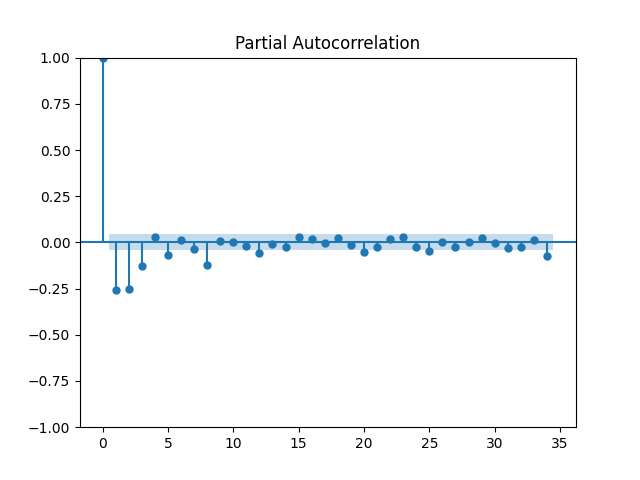} 
	\caption{\small 
		Partial autocorrelation function for the first differences of the C-Spread ($\Delta C_t$) and 5\% rejection region for the null hypothesis of no partial autocorrelation. We have statistical evidence that the first three lags are negative.
	}\label{figure:DC_PACF}
\end{figure}

When dealing with non-stationary time-series, and a cointegration relationship has been identified, the standard approach in the econometric literature is to estimate  error correction models. These models overcome the problems of using only combinations of first differences variables in the statistical estimations \citep[see e.g.,][chapter 8]{brooks2019introductory}.
In particular, we propose an error correction model of the form 
\begin{equation}
	\Delta C_t=\alpha_0+\sum_{i=1}^{3}	\beta_i\Delta C_{t-i} +\alpha_1 \Delta Z_t+\alpha_2 \Delta r_t+\alpha_3 (\psi_{t-1})+ \sum_{j=1}^{4}\delta_j Y_{j,t}+\epsilon_t\;\;, \label{eq:error_corr}
\end{equation}
where $\psi_{t}:=C_{t}-\gamma_1Z_t-\gamma_2r_t$ is the cointegration relationship, $Y_{j,t}$ are the considered control variables and $\epsilon_t$ an error term. We consider three lags for the first differences of the C-spread because the first three lags in its partial autocorrelation function are statistically  different from zero (see figure \ref{figure:DC_PACF}).

\smallskip

We estimate the error correction model  in \eqref{eq:error_corr} on the Phase III dataset  including  some additional covariates ($Y_{j,t}$) that previous contributions identified as possible drivers of the C-spread. We aim to use these additional covariates as control variables  and understand whether any of them is affecting the C-spread. As in \citet{palao2021inconvenience}, we select some broad-based financial indexes  that are usually considered as benchmarks  by financial investors worldwide: the S\&P 500 (the main equity index), the VIX (the equity volatility index), and the WTI (the reference for the crude oil price). Moreover, similarly to \citet{truck2016convenience}, we estimate a Garch(1,1) model for the conditional volatility $\sigma_t$ of the EUA spot returns. We include the estimated volatility  as an additional control variable in the error correction model. As is standard in the financial literature, for the WTI and the S\&P 500 we consider the time-series of log-returns. In table \ref{tab:correlation}, we show the Pearson correlation coefficients of the control variables.  We have statistical evidence  that the considered control variables  have low or no correlation.

\begin{table}
	\begin{center}
		\begin{tabular}{c| cc c c }  
			& $SPX_t$ &$VIX_t$&	$WTI_t$ & $\sigma_t$  \\
			\hline
			$SPX_t$	&1 & & &\\
			$VIX_t$	&	-0.16***&1&&\\
			$WTI_t$	&	0.16***	&-0.05**&1  &   \\
			$\sigma_t$	&	0.03&0.07***&0.01&1
		\end{tabular}
		\caption{\small Pearson correlation coefficients. For each coefficient, we conduct the Pearson correlation test with the null hypothesis of zero correlation. *, **, *** indicate statistical significance at the $10\%, 5\%, 1\%$ significance levels respectively. \label{tab:correlation}}
	\end{center}
\end{table}

In table \ref{tab:reg}, we report the estimated error correction models for the first differences of the C-spread $\Delta C_t$. The best-performing model in terms of BIC (and very close to the best in terms of AIC) is model (I) that considers, as independent variables, the three lags of  $\Delta C_t$ and the cointegration relationship $\psi_{t-1}$. In models (II--VI) the other variables, such as the Z-index and the risk-free rate first differences, and the control covariates (S\&P 500, VIX, WTI and $\sigma_t$) are not significant at the $1 \%$ level; instead, 
the three lags of  $\Delta C_t$ and the cointegrating relationship $\psi_{t-1}$ are  significant at $1 \%$ in all models and the estimated coefficients have always the same sign.\footnote{Results are robust to monthly and yearly fixed effects.} We observe that also  $\Delta Z_t$ is significant only at $10\%$ in models (II--VI).

\smallskip

\begin{table}[p]
	\caption{\small Error correction models for $\Delta C_t$. 
In the proposed model (I), we consider as independent variables the three lags of  $\Delta C_t$ and $\psi_{t-1}$, the cointegrating relationship at time $t-1$.
Confirming the relevance of this relation,  the Z-index and the risk-free rate first differences, 
 the S\&P 500, VIX, WTI and $\sigma_t$ are not significant at the $1\%$ level; instead, 
the three lags of $\Delta C_t$ and the cointegrating relationship $\psi_{t-1}$ are always significant at $1 \%$.
Model (I) is also the best in terms of BIC and very close to the best in terms of AIC.	
Newey-West robust standard errors appear in parentheses; *, **, *** indicate statistical significance of the parameters at the $10\%, 5\%, 1\%$ significance levels respectively.}
		
	\label{tab:reg}
	\begin{center}
\begin{tabular}{l|llllll}
	\hline
	& (I)      & (II)     & (III)  & (IV)     & (V)    & (VI)      \\
	\hline
	$\Delta C_{t-1}$        & -0.33*** & -0.33*** &        & -0.33*** &        & -0.33***  \\
	& (0.06)   & (0.06)   &        & (0.06)   &        & (0.06)    \\
	$\Delta C_{t-2}$        & -0.28*** & -0.27*** &        & -0.27*** &        & -0.27***  \\
	& (0.07)   & (0.07)   &        & (0.07)   &        & (0.07)    \\
	$\Delta C_{t-3}$        & -0.12*** & -0.12*** &        & -0.12*** &        & -0.11***  \\
	& (0.04)   & (0.04)   &        & (0.04)   &        & (0.04)    \\
	$\Delta Z_t$             &          & 0.22*    &        & 0.22*    &        & 0.24**    \\
	&          & (0.12)   &        & (0.12)   &        & (0.12)    \\
	$\Delta r_t$             &          & -0.65    &        & -0.64    &        & -0.63     \\
	&          & (0.44)   &        & (0.44)   &        & (0.44)    \\
	$\psi_{t-1}$        & -0.02*** & -0.02*** &        & -0.02*** &        & -0.02***  \\
	& (0.00)   & (0.00)   &        & (0.00)   &        & (0.00)    \\
	$WTI_t$           &          &          & -0.00  & -0.00    & -0.00  & -0.00     \\
	&          &          & (0.00) & (0.00)   & (0.00) & (0.00)    \\
	$SPX_t$           &          &          &        &          & 0.00   & -0.00     \\
	&          &          &        &          & (0.00) & (0.00)    \\
	$VIX_t$            &          &          &        &          & 0.00   & -0.00     \\
	&          &          &        &          & (0.00) & (0.00)    \\
	$\sigma_t$          &          &          &        &          & -0.00  & 0.00      \\
	&          &          &        &          & (0.00) & (0.00)    \\
	const          & 0.00     & 0.00     & -0.00  & 0.00     & -0.00  & 0.00      \\
	& (0.00)   & (0.00)   & (0.00) & (0.00)   & (0.00) & (0.00)    \\
	Obs.           & 2008     & 2008     & 2008   & 2008     & 2008   & 2008      \\
	BIC            & -20068   & -20059   & -19771 & -20052   & -19748 & -20030    \\
	AIC            & -20096   & -20098   & -19782 & -20096   & -19776 & -20092    \\
	\hline
\end{tabular}

	\end{center}
\end{table}

\smallskip

The results of the error correction model yield intriguing insights. Notably,  the financial benchmarks and the variance of the spot EUA do not impact the first differences of the C-spread. Furthermore, examining the estimated coefficients that are statistically significant in model (I), we can observe that the C-spread tends to go back to its long-run equilibrium. Let us notice that the estimated $\alpha_3$ --the coefficient of $\psi_{t-1}$-- is negative, implying a correction mechanism when the C-spread decouples from  the Z-index and the risk-free rate: e.g. when  $C_{t-1}>\gamma_1 Z_{t-1}+\gamma_2 r_{t-1}$ then $\psi_{t-1}>0$ and the C-spread tends to revert towards the cointegration relationship. Additionally, the estimated autoregressive structure with three negative lags also  indicates that, after a shock, the C-spread tends to go back to equilibrium.

\clearpage

\section{Robustness}
\label{sect:robu}

In this section, we report the robustness checks we conduct on our econometric analysis. In all cases, results are consistent with the ones presented in the previous sections.

\smallskip

We verify the robustness of the empirical results  to different definitions of the Z-index and to data-winsorization. In table \ref{tab:rob}, we report the  error correction model  \eqref{eq:error_corr} estimated when considering the different robustness checks. In all cases,  we obtain results very close to the ones presented in section \ref{sec:econometric} (cf.\ table \ref{tab:reg}). In particular, i) we compute the Z-spread of the individual issuer interpolating on the curve of the Z-spread of its active bonds at 1 year, 3 years and 5 years instead of taking the  average weighted by the bond notional (results of the error correction model are in columns I-III of the table);  ii) we compute the Z-index by weighting the Z-spreads of individual issuers with their $CO_2$ emission instead of taking the simple average (column IV of the table);
iii) we filter out possible outliers by winsoring the time-series at 95\% and at 99\% (respectively columns V and VI of table \ref{tab:rob}).
\begin{table}[H]
	\caption{\small Robusteness results for the error correction model. We consider different definition of the Z-index (columns I to IV) and data winsorization at 95\% and at 99\% (columns V and VI), observing similar results. }
	\label{tab:rob}
	\begin{center}
		\begin{tabular}{l|llllll}
			\hline
			& (I)      & (II)      & (III)      & (IV)      & (V)      & (VI)       \\
			\hline
			$\Delta C_{t-1}$        & -0.33*** & -0.33*** & -0.33*** & -0.33*** & -0.28*** & -0.34***  \\
			& (0.06)   & (0.03)   & (0.06)   & (0.06)   & (0.03)   & (0.04)    \\
				$\Delta C_{t-2}$           & -0.27*** & -0.27*** & -0.27*** & -0.28*** & -0.17*** & -0.24***  \\
			& (0.07)   & (0.07)   & (0.07)   & (0.07)   & (0.03)   & (0.04)    \\
				$\Delta C_{t-3}$           & -0.12*** & -0.12*** & -0.13*** & -0.12*** & -0.08*** & -0.12***  \\
			& (0.04)   & (0.04)   & (0.04)   & (0.04)   & (0.03)   & (0.03)    \\
			$\psi_{t-1}$        & -0.02*** & -0.02*** & -0.02*** & -0.01*** & -0.01*** & -0.02***  \\
			& (0.00)   & (0.00)   & (0.00)   & (0.00)   & (0.00)   & (0.00)    \\
			const          & 0.00     & 0.00     & 0.00     & 0.00     & 0.00     & 0.00      \\
			& (0.00)   & (0.00)   & (0.00)   & (0.00)   & (0.00)   & (0.00)    \\

			Obs.           & 2008     & 2008     & 2008     & 2008     & 2008     & 2008      \\
			\hline
		\end{tabular}
	\end{center}
\end{table}

\smallskip

\begin{figure}
	\centering
	\includegraphics[width=0.75\textwidth]{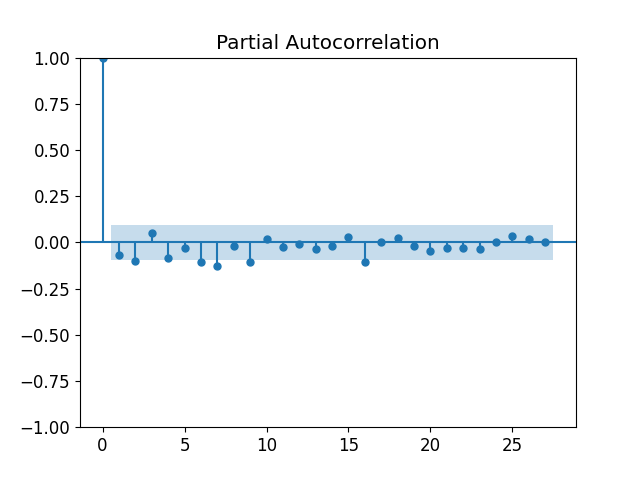} 
	\caption{\small 
		Partial autocorrelation function for the first differences of the C-Spread for weekly data and 5\% rejection region for the null hypothesis of no partial autocorrelation.
	}\label{figure:DC_PACF_week}
\end{figure}

Repeating the analysis considering only weekly data, we obtain some interesting insights that deserve additional comments. In figure \ref{figure:DC_PACF_week},  we report the	partial autocorrelation function for the first differences of the C-Spread for aggregated weekly data. 
In the daily  case, we have shown that the first three lags of the partial autocorrelation are significant at the 5\% level. As expected, in the weekly case, we observe that no partial autocorrelation lag is significant at 5\%.
In table \ref{tab:Models_week}, we report the estimated error correction models for $\Delta C_t$ considering  weekly data. In this case, consistently with the partial autocorrelation function, the best-performing model  in terms of BIC and AIC is model (I) that depends only from 	the cointegration relationship $\psi_{t-1}$ and not from past differences. This is another indication that the C-spread tends to revert to the credit spread in the long-run while the autoregressive structure is only relevant in the short-run.
	
	\begin{table}
		\caption{\small Error correction models for $\Delta C_t$ considering weekly data. The cointegration relationship remains significant at 1\% in all models. }
		\label{tab:Models_week}
		\begin{center}
			\begin{tabular}{l|lll}
				\hline
				& (I)    & (II)      & (III)      \\
				\hline
				$\Delta C_{t-1}$        &          & -0.03    & -0.04     \\
				&          & (0.08)   & (0.08)    \\
				$\Delta C_{t-2}$        &          & -0.07    & -0.07     \\
				&          & (0.07)   & (0.07)    \\
				$\Delta C_{t-3}$        &          & 0.08     & 0.08      \\
				&          & (0.06)   & (0.05)    \\
				$\Delta Z_t$             &          &          & 0.91      \\
				&          &          & (0.57)    \\
				$\Delta r_t$             &          &          & 0.15      \\
				&          &          & (0.90)    \\
				$\psi_{t-1}$        & -0.03*** & -0.03*** & -0.03***  \\
				& (0.01)   & (0.01)   & (0.01)    \\
				const          & -0.00    & -0.00    & -0.00     \\
				& (0.00)   & (0.00)   & (0.00)    \\

				Obs.           & 417      & 417      & 417       \\
				BIC            & -4080    & -4067    & -4057     \\
				AIC            & -4088    & -4087    & -4086     \\
				\hline
			\end{tabular}
		\end{center}
	\end{table}

\newpage
\section{Conclusions and policy implications}
\label{sec:conclusions}

Understanding the relationship between spot and futures prices is of crucial importance for all participants in the carbon market. The carbon trading scheme works only if markets for carbon provide enough liquidity and pricing accuracy. The efficiency of the EU-ETS market is particularly important for emission-intensive firms, policymakers, and all EU citizens.
If these contracts are efficiently priced,  participating countries and covered installations can achieve environmental compliance in a cost-effective and optimal manner \citep[see e.g.,][]{krishnamurti2011efficiency}.

\bigskip

We have described in detail the relationship between the spot and futures EUA market that is in line with a market where the majority of participants cannot borrow at the risk-free rate, but need to consider a credit spread in their cost of funding. 
The cost of storing allowances is negligible, and there is no obvious benefit to hold spot allowances as there is for physical commodities. Therefore, the C-spread (or the negative convenience yield) should be equal to zero as in any financial asset paying no dividend.\footnote{The only (negative) convenience yield could be related to an unlikely scenario where regulators might change the bankability of allowances either forbidding the use of allowance vintages past a certain number of years or introducing a new tax on banked allowances.} For this reason, multiple studies have tried to understand why this C-spread is persistently above zero.

\smallskip
In this study, we have solved the EU-ETS market C-spread puzzle by showing that this spread is caused by the credit spread of compliance entities.  This link is due to the fact that the financial decisions of a compliance entity in the scheme  are based on the comparision of C-spread  and its credit spread.  Our claims are consistently supported by the empirical results in section \ref{sec:econometric}. In particular, we obtain three  key results  in the empirical analysis.

\smallskip

First, we analyze in detail the volume of the EU-ETS spot and futures market and identify the most liquid contracts that we can use to build the time-series of C-spread. We also construct the Z-index, our proxy for the credit spread of market participants in the EU-ETS.

\smallskip

Secondly, we are the first to identify a cointegration relationship for the C-spread. We present statistical evidence, based on the GLS ADF test and on the \citet{johansen1988statistical} procedure, that the C-spread, the Z-index, and the three-month risk-free rate are integrated of order one and cointegrated (cf.\ table \ref{tab:coint}). The most relevant component of the C-spread  is explained by the credit spread  and the cointegration  coefficients are shown in (\ref{eq:coint}).

\smallskip

Lastly, we  estimate the error correction model  \eqref{eq:error_corr} for the first differences of the C-spread. We observe that the  financial indexes that were identified in the literature as possible drivers of the C-spread and the variance of the EUA spot do not play any role once the cointegration relationship has been taken into account. Moreover, we  point out  that, in the error correction framework, the estimated coefficients confirm that the C-spread tends to revert to the credit spread (proxied by the Z-index) in the long-run.

\bigskip

As pointed out by \citet{charles2013market}, the presence of the positive C-spread in the cost-of-carry is an indicator of inefficiency in the carbon markets.
If markets are inefficient there is a greater role to be played by
 regulators to improve information flows and reduce 
manipulations \citep{stout1995stock}. It is imperative that policymakers
address the issue of the positive C-spread because failure to tackle EUA market inefficiencies  affects  EU-ETS credibility, endangering the European Union decarbonization policy and its energy transition.
 If the EU-ETS compliance entities lack foresight and are not convinced by the long term credibility of the scheme, they are unlikely to anticipate EUA scarcity. Consequently, they will myopically lack the incentive to buy EUA in advance ensuring that market prices reflect the future carbon emission constraining caps. Thus, this lack of foresight will not encourage compliance entities  to decrease emissions in the short term and invest in long-term decarbonization projects  \citep{bocklet2020does,pietzcker2023eu}.

The main policy implication of our study is that the simplest solution for dealing with the market inefficiency identified by the C-spread --thus enhancing the foresight of  compliance entities-- involves  including the EUA in the list of eligible collaterals for Euro-system credit operations.  Designating the EUA as high-quality collateral  would engender the formation of a EUA repo market and close the C-spread, improving market liquidity \citep{mesonnier2017interest,cuzzola2023stress}.

Including the EUA in the roster of eligible Euro-system collaterals confers also other benefits to the green transition: it will not only reinforce the  credibility of the EU-ETS but also signal a long-term committed stance by policymakers toward sustaining  the scheme in the foreseeable future. 
Numerous studies have advocated the introduction of a price floor to the EU-ETS framework, underscoring concerns about avoiding that the price drops to zero as it happened at the end of Phase I \citep[see e.g.,][]{flachsland2020avoid,bolat2023there}. The exigency of implementing such a radical change to the scheme is considerably diminished in a more efficient, liquid and credible EUA market.
	
	\section*{Acknowledgments}
	The authors are grateful to Rüdiger Kiesel and Paul Tautorat for some interesting preliminary discussion on the EU-ETS market.
	\bibliography{sources}
	\bibliographystyle{tandfx}


\end{document}